\documentclass[twocolumn,showpacs,preprintnumbers,nofootinbib,prd,superscriptaddress]{revtex4-1}
\usepackage{graphicx,amssymb,amsmath,amsthm,amsfonts,epsfig}

\usepackage[linktocpage]{hyperref}
\usepackage[usenames]{color}
\usepackage{epstopdf}

\def\nn{\nonumber}

\newcommand{\ben}{\begin{enumerate}}
\newcommand{\een}{\end{enumerate}}

\def\be{\begin{equation}}
\def\ee{\end{equation}}
\def\bea{\begin{eqnarray}}
\def\eea{\end{eqnarray}}
\newcommand{\beq}{\begin{eqnarray}}
\newcommand{\eeq}{\end{eqnarray}} 

\begin{document}
\title{Matter around Kerr black holes in scalar-tensor theories: \\
scalarization and superradiant instability}

\author{Vitor Cardoso} 
\affiliation{CENTRA, Departamento de F\'{\i}sica, Instituto Superior T\'ecnico, Universidade T\'ecnica de Lisboa - UTL,
Av.~Rovisco Pais 1, 1049 Lisboa, Portugal.}
\affiliation{Perimeter Institute for Theoretical Physics
Waterloo, Ontario N2J 2W9, Canada.}
\affiliation{Department of Physics and Astronomy, The University of Mississippi, University, MS 38677, USA.}
\affiliation{Faculdade de F\'{\i}sica, Universidade 
Federal do Par\'a, 66075-110, Bel\'em, Par\'a, Brazil.}

\author{Isabella P. Carucci}
\affiliation{CENTRA, Departamento de F\'{\i}sica, Instituto Superior T\'ecnico, Universidade T\'ecnica de Lisboa - UTL,
Av.~Rovisco Pais 1, 1049 Lisboa, Portugal.}
\affiliation{Dark Cosmology Centre, Niels Bohr Institute, University of Copenhagen, Juliane Maries Vej 30, 2100 Copenhagen, Denmark}

\author{Paolo Pani}
\affiliation{CENTRA, Departamento de F\'{\i}sica, Instituto Superior T\'ecnico, Universidade T\'ecnica de Lisboa - UTL,
Av.~Rovisco Pais 1, 1049 Lisboa, Portugal.}
\affiliation{Institute for Theory $\&$ Computation, Harvard-Smithsonian
CfA, 60 Garden Street, Cambridge, MA, USA.}

\author{Thomas P. Sotiriou}
\affiliation{SISSA, Via Bonomea 265, 34136, Trieste, Italy and INFN, Sezione di Trieste, Italy.}

\date{\today} 

\begin{abstract} 
In electrovacuum stationary, asymptotically flat black holes in scalar-tensor theories of gravity are described by the Kerr-Newman family of solutions, just as in general relativity.
We show that there exist two mechanisms which can render Kerr black holes unstable when matter is present in the vicinity of the black hole, as this induces an effective mass for the scalar. The first mechanism is a tachyonic instability that appears when the effective mass squared is negative, triggering the development of scalar hair --- a black hole version of ``spontaneous scalarization''. The second instability is associated with superradiance and is present when the effective mass squared is positive and when the black hole spin exceeds a certain threshold. The second mechanism is also responsible for a resonant effect in the superradiant scattering of scalar waves, with amplification factors as large as $10^5$ or more.
\end{abstract}

\pacs{
04.50.Kd
04.70.-s
04.25.Nx
}
\maketitle

\section{Introduction}
Scalar-tensor theories of gravity are described by the action~\cite{Fujii:2003pa,valeriobook}
\bea
\label{actionST}
S&=&\frac{1}{16\pi G}\int d^4x \sqrt{-g}\left(F(\phi)R-Z(\phi)g^{\mu\nu}\partial_{\mu}\phi\partial_{\nu}\phi-U(\phi)\right)\nonumber\\
&+&S(\Psi_m;g_{\mu\nu})\,,
\eea
where $R$ is the Ricci scalar of the spacetime metric $g_{\mu\nu}$, $\phi$ is a scalar field, $\Psi_m$ collectively denotes the matter fields which are minimally coupled to $g_{\mu\nu}$ and $G$ is a constant. Choosing the functions $F$, $Z$ and $U$ determines a specific theory within the class, up to a degeneracy due to the freedom to redefine the scalar.\footnote{{\em e.g.}~$F(\phi)$ can be set equal to $\phi$ without loss of generality, see for instance Ref.~\cite{Sotiriou:2007zu}; however, we find it convenient to avoid this redefinition here.} Theories with a scalar field nonminimally coupled to the metric have been studied extensively and action (\ref{actionST}) describes a fairly (though not the most) general class of such theories.

Brans--Dicke theory \cite{bd} is perhaps the most well-known scalar-tensor theory and it corresponds to the choice $F=\phi$, $Z=\omega_0/\phi$, $U=0$. Scalar-tensor theories can be seen as generalizations of Brans--Dike theory, and consequently, as more general theories with a varying gravitational coupling. They have also been considered extensively in cosmology, as a rather general parametrization for dark energy~\cite{Clifton:2011jh}. Action (\ref{actionST}) can be thought of as a low-energy effective action of a more fundamental theory, which would come to specify $F$, $Z$ and $U$. For example, bosonic string theory leads to $F=\phi$, $Z=-\phi^{-1}$, $U=0$. In this spirit, studying scalar-tensor theory may provide a glimpse into the phenomenology of quantum gravity candidates with extra scalar degrees of freedom.

If there is a scalar field mediating the gravitational interaction, the obvious question that arises is why there is not evidence for it in local tests of gravity. Given that the range of the associated interaction is inversely proportional to the mass, an obvious answer would be that the scalar is sufficiently massive to avoid detection so far. Such a simple resolution is phenomenologically perfectly acceptable as long as one is not expecting the scalar to play any role at large scales. For the scalar to be relevant in late time cosmology --- {\em e.g.}~if it is to account for dark energy --- it would have to mediate a long range interaction. Various screening mechanisms have been devised, that allow the scalar to do so and still evade solar system and local gravity tests \cite{Khoury:2003aq,Hinterbichler:2010es}.

This highlights the need to consider the phenomenology of scalar-tensor theory in the strong gravity regime. It is well known that  important constraints can be obtained from compact stars and black holes~\cite{Will:2005va}. Moreover, understanding the role of a scalar field and the effect it might have in the structure of these objects is interesting in its own right. 

Intuitively, black holes are much simpler objects to study than compact stars, as they are vacuum solutions. One might worry in fact that there might be too simple to be interesting:
It has been shown in Ref.~\cite{Sotiriou:2011dz} that, once one assumes stationarity and asymptotic flatness, the scalar field around a black hole settles to a constant and the metric satisfies Einstein's equations. Therefore, black holes that are endpoints of gravitational collapse and that are isolated will have to be described by the Kerr-Newman family of solutions, as in general relativity (GR).  

Nonetheless, even these black holes, identical as they may be to those of GR, can still be powerful probes of the existence of a scalar field. This is because, once perturbed, they will behave differently than their GR counterparts  and this can be used to detect deviations from Einstein's theory \cite{Barausse:2008xv}. In particular, it has been shown that remarkable effects, such as floating orbits for test particles, can take place in the vicinity of Kerr black holes in scalar-tensor theory \cite{Cardoso:2011xi,Yunes:2011aa}. These effects are associated with the existence of a scalar mode in the spectrum of perturbations. 

Generically the equations of scalar-tensors theories require the scalar to have a nontrivial profile whenever matter is present. That indicates that black holes with matter in their vicinity, {\em e.g.}~due to the presence of an accretion disk, will tend to have scalar hair. It is, therefore, an interesting open question to determine whether this has an observable signature when it comes to realistic, astrophysical black holes. Nonetheless, as we discuss in more detail below, there exists a specific subclass of theories that allows for constant $\phi$ solutions even in the presence of matter. In these theories, one could entertain the thought that the spacetime around compact stars and black holes with matter in their vicinity could be identical to that for the same configuration in GR.

Compact stars have been studied extensively in this subclass. It has been shown that, even though a GR solution with a constant scalar field is admissible in general, it is not always the energetically preferred solution. So long as the central density of the star remains below a certain threshold, a ``hairless'' configuration is indeed preferred. Once this threshold density is exceeded, however, the scalar develops a non-trivial profile and the spacetime deviates from its GR counterpart. This unexpected effect is referred to  as ``spontaneous scalarization" \cite{Damour:1993hw,Damour:1996ke,Harada:1997mr,Pani:2010vc}. Remarkably, it can lead to large deviations from GR in theories which have identical phenomenology to the latter in the solar system or local gravity tests.

We show here that ``spontaneous scalarization'' is not restricted to stars. It can also take place when black holes are surrounded by a sufficient amount of matter. At the perturbative level, the presence of this matter induces a negative effective mass squared for the scalar and spontaneous scalarization manifests as a tachyonic instability. As a consequence, the black hole is forced to develop scalar hair. When the effective mass squared of the scalar is positive spontaneous scalarization does not occur. However, another type of instability can occur, which is driven by superradiance~\cite{Teukolsky:1974yv,Press:1972zz}. As we show below, the same ``spontaneous superradiance" effect is responsible for a resonant amplification in the scattering of scalar waves off a spinning black hole.

At first sight the subclass of theories in which the phenomena discussed above take place might seem quite restricted. However, closer analysis shows that, once certain viability criteria are taken into account, the theories in question, or at least their specific features that are associated with the phenomenology discussed here, turn out to be much more generic than one would have thought. Hence, these phenomena have a good chance of being physically relevant and could in principle be used to constrain scalar-tensor theories to unprecedented levels. This will be discussed in more detail in the next sections.

We use the signature $(-,+,+,+)$ for the metric and adopt natural units $\hbar=c=G=1$.

\section{Framework}
Let us start with the action (\ref{actionST}) of a generic scalar-tensor theory in the Jordan frame~\cite{Fujii:2003pa}.
By performing the following transformations
\begin{eqnarray}
 g^E_{\mu\nu}&=&F(\phi)g_{\mu\nu}\,,\label{metricEinstein}\\
 \Phi(\phi)&=&\frac{1}{\sqrt{4\pi}}\int d\phi\,\left[\frac{3}{4}\frac{F'(\phi)^2}{F(\phi)^2}+\frac{1}{2}\frac{Z(\phi)}{F(\phi)}\right]^{1/2}\,,\\
 A(\Phi)&=&F^{-1/2}(\phi)\,,\\
 V(\Phi)&=&\frac{U(\phi)}{F^2(\phi)}\,,
\end{eqnarray}
the theory can can be written in the Einstein frame as
\bea
S&=&\int d^4x \sqrt{-g^E}\left(\frac{R^E}{16\pi}-\frac{1}{2}g^E_{\mu\nu}\partial^\mu\Phi\partial^\nu\Phi-\frac{V(\Phi)}{16\pi}\right)\nonumber\\
&+&S(\Psi_m;A(\Phi)^2g_{\mu\nu}^E)\,.
\eea
In the Einstein frame the scalar field is minimally coupled to gravity, but any matter field $\Psi_m$ is coupled to the metric $A(\Phi)^2g_{\mu\nu}^E$.
The field equations in the Einstein frame read
\begin{eqnarray}
G_{\mu\nu}^E&=&8\pi 
\left(T_{\mu\nu}^E+\partial_\mu\Phi\partial_\nu\Phi-\frac{g_{\mu\nu}^E
}{2}(\partial\Phi)^2\right)-\frac{g_{\mu\nu}^E}{2} 
V(\Phi)\,,\nn\\\label{einein}\\\label{einscalar}
\square^E\Phi&=&-\frac{A'(\Phi)}{A(\Phi)}T^E+\frac{V'(\Phi)}{16\pi}\,,
\end{eqnarray}
where the stress-energy tensor in the Einstein frame is related to the physical one by
\begin{equation}
 {T^\mu_\nu}^E=A^4(\Phi) T^{\mu}_{\nu}\,,\quad T_{\mu\nu}^E=A^2(\Phi) T_{\mu\nu}\,,\quad T^E=A^4(\Phi) T\,.\nn
\end{equation}
%
Let us now assume a general analytical behavior of the potentials around $\Phi\sim \Phi^{(0)}$:
\begin{eqnarray}
 V(\Phi)=\sum_{n=0}V_n(\Phi-\Phi^{(0)})^n\,,\\
 A(\Phi)=\sum_{n=0}A_n(\Phi-\Phi^{(0)})^n\,.\label{expansionA}
\end{eqnarray}
Expanding the field equations to first order in $\varphi\equiv\Phi-\Phi^{(0)}\ll1$, we obtain~\cite{Yunes:2011aa}
\begin{eqnarray}
&&G_{\mu\nu}^E=8\pi \left(T_{\mu\nu}^E+\partial_\mu\Phi_0\partial_\nu\Phi_0-\frac{g_{\mu\nu}^E}{2}(\partial\Phi_0)^2\right)-
\frac{g_{\mu\nu}^E}{2} V_0\nonumber\\
&&+8\pi\left(\partial_\mu\Phi_0\partial_\nu\varphi+\partial_\mu\varphi\partial_\nu\Phi_0-g_{\mu\nu}^E\partial_\mu\Phi_0\partial^{\mu}\varphi\right)-\frac{g_{\mu\nu}^E}{2} V_1\varphi\,,\nonumber\\ \label{Einsteinlin} \\
&&\square^E\Phi_0+\square^E\varphi=-\frac{A_1}{A_0}T^E+\frac{V_1}{16\pi} +\frac{V_2 \varphi}{8\pi}\nn\\
&&+\varphi T^E\left(\frac{A_1^2}{A_0^2}-2\frac{A_2}{A_0}\right)\,, \label{KGlin}
\end{eqnarray}
The term $V_0$ is related to a cosmological constant. For simplicity, we focus on asymptotically flat solutions, although we expect most of our results to be valid also in asymptotically de Sitter backgrounds. We then set $V_0=0=V_1$. From Eq.~\eqref{KGlin} it is clear that in presence of matter $\Phi_0={\rm const}$ does not solve the background equations unless $A_1=0$. Thus, if $A_1\neq0$, any black hole solution surrounded by matter -- if it exists -- will necessarily be endowed by a nontrivial scalar profile. These configurations are different from their GR counterparts and the strength of the deviations depends on the coupling constant $A_1$ and on the energy content of the surrounding matter.

On the other hand, if $A_1=0$ the field equations above admit any solution of Einstein's equations with a constant background scalar field. Our goal here is to show that, nonetheless, perturbative effects might give rise to novel mechanisms that force the hairless black hole to develop a nontrivial scalar profile.
Let us then set $A_1=0$ and consider a background GR solution, which solves the lowest-order equations above. All that remains, to first order in $\varphi$, is the Klein-Gordon equation
\beq
\left[\square^E-\frac{V_2}{8\pi}+\frac{2A_2}{A_0} T^E\right]\varphi&\equiv& \left[\square^E-\mu_s^2(r)\right]\varphi=0\,,\\
\mu_s^2(r)&\equiv& \frac{V_2}{8\pi}-\frac{2A_2}{A_0} T^E\,. \label{effectivemass}
\eeq
%
Thus, couplings of scalar fields to matter are equivalent to an effective radius-dependent mass for the scalar field. The term $V_2$ is related to a standard mass term.  Its presence is an unnecessary complication that does not change our discussion qualitatively. We therefore neglect this term in the following. 
On the other hand, depending on the sign of $A_2$ and and $T^E$, the effective mass squared can be either positive or negative. Depending on the sign, two types of instabilities may drive the background solution to develop scalar hair; we discuss them separately in the next sections.

Before moving on, a comment is due on how restrictive our essential assumptions are, {\em i.e.}~the fact that we require $A(\Phi)$  to be analytic around $\Phi^{(0)}$ and to have $A_1=0$, that is, to have an extremum for this value. In fact, for $\Phi^{(0)}=$constant to be a solution at the non-perturbative level, one would need $A'(\Phi^{(0)})=0$. Transforming back to the Jordan frame, that would translate to $Z(\phi_0)\to \infty$, with $\phi_0$ being the corresponding value of $\phi$ in this solution. So, one might be tempted to think that theories with such characteristic are unphysical or non-representative. 

However, this does not really seem to be the case. As is obvious from 
Eq.~(\ref{einscalar}), $\alpha\equiv A'/A$ controls the effective 
coupling between the scalar and matter. Various observations, such as 
weak gravity constraints and test for violations of the strong 
equivalence principle, seem to require $\alpha$ to be negligibly small 
when the scalar takes its asymptotic value 
\cite{Damour:1998jk,Damour:1996ke,Freire:2012mg}.  This implies that a 
configuration in which the scalar is constant and $\alpha\approx 0$ is 
most likely to be at least an approximate solution in most viable 
scalar-tensor theories. Such a configuration would satisfy our 
assumptions to adequate precision, and in this spirit our 
analysis and results appear to be rather generic when one restricts 
ones attention to viable scalar-tensor theories. In a certain sense, 
we are choosing to look at the special subclass of theories in which 
the effects in question are more prominent and easier to calculate, 
but that does not mean that they will not  be generic to viable 
scalar-tensor theories.

\section{Spontaneous scalarization}
The most important result of our analysis is that adding a matter distribution $T^E$ around black holes forces the scalar field to be spontaneously excited and develop a non-trivial configuration. In other words, even though GR (with constant scalar field) is a solution of the field equations, it is not the entropically
preferred configuration. 
This phenomenon is the direct analog of spontaneous scalarization 
first discussed for compact stars by Damour and Esposito-Far\`ese and 
many others since 
\cite{Damour:1993hw,Damour:1996ke,Harada:1997mr,Pani:2010vc,Barausse:2012da}. At 
linear level, spontaneous scalarization manifests itself as a 
tachyonic instability triggered by a \emph{negative} effective mass 
squared. 

Let us first consider for simplicity the case in which $T^E$ is spherically symmetric, $T^E=T^E(r)$, and its backreaction in the geometry is negligible. This  assumption will be dropped later on. In this probe limit the background metric solution is a Schwarzschild black hole. After a decomposition in spherical harmonics $\varphi(t,r,\theta,\phi)=\sum_{lm}\frac{\Psi_{lm}(r)}{r}e^{-i\omega t}Y_{lm}(\theta,\phi)$, the scalar field then obeys the equation
\beq
&&\frac{d^2\Psi(r)}{dr_*^2}+\left[\omega^2-{\cal V}(r)\right]\Psi(r)=0\,,\label{KGsphe}\\
&&{\cal V}(r)=f\left(\frac{l(l+1)}{r^2}+\frac{2M}{r^3}+\mu_s^2(r)\right)\,,
\eeq
where $f=1-2M/r$ and the tortoise coordinate is defined as $dr/dr_*=f$. It is understood that the wavefunction $\Psi$ carries an $lm$ subscript which we omit for compactness of notation from here onwards. This equation
can be thought of as an eigenvalue equation for the possible characteristic frequencies $\omega$, when the 
eigenfunctions $\Psi(r)$ are required to satisfy appropriate boundary conditions, viz. out-going waves
at spatial infinity, $\Psi \sim e^{+i\omega r_*}$ and in-going at the horizon, $\Psi \sim e^{-i\omega r_*}$. 
These conditions are also equivalent to causal behavior of compact-domain perturbations and regularity, respectively.
Given the time-dependence of $\varphi$, an unstable mode corresponds to a characteristic frequency with a positive imaginary component.
But this also means that the mode is spatially bounded (decaying exponentially at spatial infinity and at the black hole horizon).
In this case, one can make contact with and borrow some powerful results from quantum mechanics. In particular,
a {\it sufficient} condition for this potential to lead to an instability is that \cite{Buell}
\be
\int_{2M}^{\infty} \frac{{\cal V}}{f}dr<0\,, \label{sufficient}
\ee
which yields the instability criterion
\be
2\frac{A_2}{A_0}\int_{2M}^{\infty} T^E dr>\frac{2l(l+1)+1}{4M}\,.
\ee
The above is a very generic, analytic result. We have checked numerically for some specific cases discussed  below that the inequality is nearly saturated for all interesting matter configurations. Let us consider some simple toy models for demonstrative purposes.

\noindent {\bf{\em Model I.}} 
\be
\mu_s^2\equiv -2\frac{A_2}{A_0} T^E=-\Theta(r-r_0)\beta M^{n-3}\frac{r-r_0}{r^n}\,.
\ee
Because $T^E\sim -\rho$, we get
\be
\beta = -\frac{A_2}{2 \pi A_0}\frac{\mu}{M}(n-4)(n-3)\left(\frac{r_0}{M}\right)^{n-4}\,,
\ee
where $\mu$ is the mass of the spherical distribution and its finiteness requires $n>4$.
This matter distribution, chosen quite arbitrarily to make our point, has two important features: it models the existence of an innermost stable circular orbit close
to the event horizon, by not allowing matter to be closer than $r=r_0$, and it decays to zero at large distances provided $n$ is large enough. The distribution of matter $T^{E}$ is zero for $r<r_0$
and peaks at $r=n/(n-1)\,r_0$, after which it decays like a power-law $r^{-n+1}$ at large distances. We get that spontaneous scalarization occurs for
\be
\beta \gtrsim \frac{2l(l+1)+1}{4}(n-2)(n-1)\left(\frac{r_0}{M}\right)^{n-2}\,,
\ee
or equivalently, the more transparent bound
\be
-\frac{A_2}{A_0}\frac{\mu}{M} \gtrsim 2\pi \frac{2l(l+1)+1}{2}\frac{(n-2)(n-1)}{(n-4)(n-3)}\left(\frac{r_0}{M}\right)^ 2\,.\label{model1}
\ee

\noindent {\bf{\em Model II.}}
\be
\mu_s^2\equiv -2\frac{A_2}{A_0} T^E=-\frac{\beta}{M^2} \left(\Theta(r-r_0)-\Theta(r-r_0-L)\right)\,.
\ee
This represents a shell of constant density, with inner and outer radius $r_0$ and $r_0+L$ respectively.
For this model, spontaneous scalarization occurs for
\be
\beta \gtrsim \frac{2l(l+1)+1}{4}\frac{M}{L}\,,
\ee
or equivalently, 
\be
-\frac{A_2}{A_0}\frac{\mu}{M} \gtrsim \pi 
\frac{2l(l+1)+1}{6}\frac{3 r_0 
(r_0+L)+L^2}{M^2}\,.\label{model2}
\ee
%
%
These results show that a minimum mass $\mu$ is necessary in order for 
spontaneous scalarization to occur. 
Binary pulsar experiments 
constrain $A_2/A_0\gtrsim-26$~\cite{Damour:1996ke}. Using the maximum 
allowed value we get 
$\mu/M\gtrsim0.1 (r_0/M)^2$, for $l=0$ and in the $n\gg1$ and $L\ll 
r_0$ limits of Eqs.~\eqref{model1} and~\eqref{model2}, respectively.
From a dynamical perspective, we could envisage accreting black holes 
for which the accreting mass is small enough that they are still 
characterized by GR. These GR solutions would be spontaneously 
scalarized as soon as the accretion mass rises up above the threshold 
mass.
Finally, note that it is the combination $A_2 T^E$ that regulates 
the instability. If some exotic form of matter surrounds the black 
hole such that $T^E>0$, then the instability occurs for 
\emph{positive} values of $A_2/A_0$, which are not constrained by 
observations.

\noindent {\bf{\em Model III.}}
The results above were obtained assuming the background metric is that of a Schwarzschild black hole even in the presence of matter. For consistency, this requires $\mu\ll M$ and it might seem hard to stay within the range of validity of this approximations and satisfy the inequalities in Eqs.~(\ref{model1}) and (\ref{model2}). However, as might be anticipated, the instability is quite generic and occurs also for consistent background solutions, as we now show. Let us consider a spherically symmetric black hole endowed with a spherical thin-shell located at some radius $R$. This is an exact solution of Einstein's equation and it is discussed in the next section below. The background metric reads:
\begin{equation}
 ds^2=-(1-2m(r)/r)dt^2+(1-2m(r)/r)^{-1}dr^2+r^2 d\Omega^2\,,\nn
\end{equation}
where $m(r)=M$ for $r>R$ and $m(r)=M_{\rm int}$ for $r<R$. The surface stress-energy tensor of the shell is given below and, once the surface energy density $\sigma$ and pressure $P$ are specified, Israel's junction conditions~\cite{Israel:1966rt} provide the internal mass $M_{\rm int}$ and the shell location $R$ in terms of $\sigma$, $P$ and $M$. In this case, the sufficient condition~\eqref{sufficient} becomes:
\begin{equation}
 \frac{2 A_2}{A_0}(2P-\sigma)>\frac{2l (l+1)+1}{4 M_{\rm int}}+\frac{M-M_{\rm int}}{R^2}>0\,,
\end{equation}
which reduces to the case discussed above when $M_{\rm int}=M$.
Therefore, if $\sigma>2P$ scalarization may occur if $A_2$ is sufficiently negative, whereas if $\sigma<2P$ the instability occurs for large enough values of $A_2>0$. Although not strictly necessary, it is customary to impose the following energy conditions: $\sigma\geq0 ,\ \sigma\geq |P| $ and $ \sigma+2P\geq 0$.

The above models all consider spherically symmetric matter distribution, we now show that spontaneous scalarization is in fact
very generic. Consider a generic matter distribution
such as a thin accretion disk, modelled by $\mu_s^2=\mu_s^2(r,\theta,\phi)=\sum_l \mu_{s\,l}^2(r)Y_{lm}(\theta,\phi)$.
We drop the argument of $\mu_{s\,l}^2(r)$, with the understanding it is a function of radial distance alone.
Then the Klein-Gordon equation is reduced to
\beq
&&\sum_l \left\{\frac{d^2\Psi_l(r)}{dr_*^2}+\left[\omega^2-f\left(\frac{l(l+1)}{r^2}+\frac{2M}{r^3}\right)\right]\Psi_l(r)\right\} Y_{lm}\nonumber\\
&&=\sum_l \sum_{l'}f\mu_{s\,l'}^2\Psi_lY_{l'm'}Y_{lm}\,.
\eeq
Multiplying by $Y_{00}$, integrating on the two-sphere and using the orthogonality properties of spherical harmonics, we get
\be
\frac{d^2\Psi_0(r)}{dr_*^2}+\left[\omega^2-f\left(\frac{2M}{r^3}+\frac{\mu_{s,\,0}^2}{\sqrt{4\pi}}\right)\right]\Psi_0(r)=0\,.
\ee
Thus, the $l=0$ component of the scalar field is always described by an equation similar to Eq.~\eqref{KGsphe}, which is completely decoupled from the higher harmonics. Because in general (but not always) $\mu_{s,\,0}^2\neq 0$,
we conclude that scalarization must occur \emph{at least} at the level of the $l=0$ mode. We focused on the $l=0$ mode for simplicity, but it is most likely that higher $l$ modes may be unstable even when $\mu_{s,\,0}^2= 0$.
Finally, using the numerical methods of Section~\ref{sec:superradiant} one can show that
spontaneous scalarization is active also when the black hole rotates.

\subsection{The final state of spontaneous scalarization}
%
\begin{figure}
\begin{center}
\begin{tabular}{c}
\epsfig{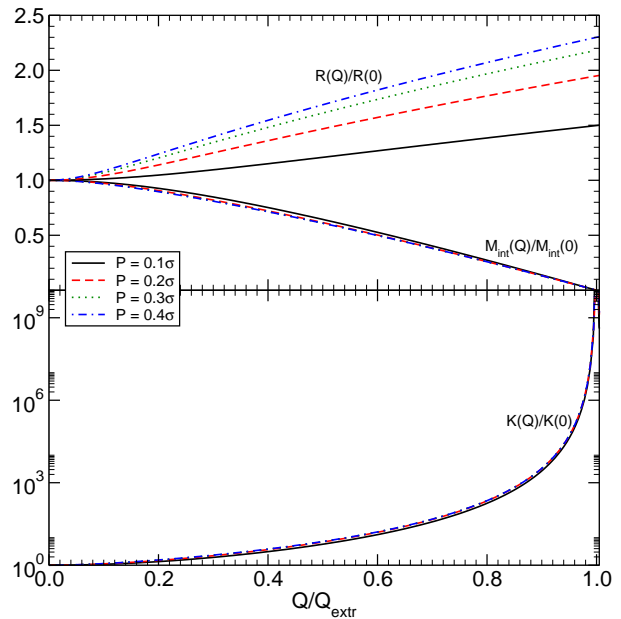}
\end{tabular}
\end{center}
\caption{\label{fig:shell}
The internal mass, $M_{\rm int}\equiv M-C/2$, the shell radius $R$ and the Kretschmann scalar $K=R_{abcd}R^{abcd}$ at the black-hole radius for a hairy black hole as functions of the scalar charge $Q$ and for different values of the shell pressure $P$ at fixed total mass $M$. The vertical axis the corresponding value in GR, i.e. $Q=0$. The horizontal axis is normalized by the charge $Q_{\rm extr}$ corresponding to the extremal solution with $M_{\rm int}=0$. With these normalizations, the quantities $M_{\rm int}$ and $K$ are almost universal, whereas the shell radius $R$ is not.
We show the exact solutions computed numerically, but the perturbative solutions discussed in the text agree very well with the exact result~\cite{letter}. In this example we set $\sigma M=10^{-3}$. Similar results hold for different values of the shell energy density.
}
\end{figure}
We showed that, for generic scalar-tensor theories, GR solutions
can be unstable in the presence of matter. To understand the development of such instability and the approach to the final state, a nonlinear
time evolution is mandatory. However, if one is interested in the final state only, interesting information
can be obtained by looking only at stationary solutions of the field equations with the same symmetries.
Let us now focus on the final state of spherically symmetric configurations, for which Schwarzschild is a background solution,
and let us work out the spontaneous scalarization final state for a simple spherically symmetric shell of matter.
Spacetime is described by
\be
ds^2=-h(r)dt^2+f(r)^{-1}dr^2+r^2d\Omega^2
\ee
Because we are considering a zero-thickness shell, the matter content is zero everywhere and Klein-Gordon equation can be integrated to yield
(we work with the original, full scalar field $\Phi$):
\be
\Phi'=\frac{Q}{r^2\sqrt{fh}}\,.
\ee
The scalar charge $Q$ can be determined as a function of the matter density $\sigma$ and pressure $P$ on the shell. We will come to this point shortly.
This equation is interesting, because it implies that if there is an horizon ($f=0$) inside the shell, the requirement that the 
scalar field be regular implies immediately that $\Phi={\rm const}$ inside the shell. Thus, we get a Schwarzschild interior.
Furthermore, the $tt$ and $rr$ components of Eq.~(\ref{einein}) yield
\beq
4\pi Q^2+r^2 h\left(f+r f'-1\right)&=&0\,,\label{eqE1}\\
4\pi Q^2+r^2 h\left(1-f\right)-r^3 f h'&=&0\,.\label{eqE2}
\eeq
Once \eqref{eqE1}-\eqref{eqE2} are satisfied, the $\theta \theta$ and $\phi \phi$ are trivially satisfied as well.
For a shell made of a layer of perfect fluid, the surface stress-energy tensor reads
\be
S_{ab}^E=\sigma u_au_b+P(\gamma_{ab}+u_au_b)\,,\label{Sab}
\ee
where $P,\sigma$ denote pressure and density, $\gamma_{ab}$ is the induced metric, $u_a$ is the on-shell
four-velocity and all quantities refer to the Einstein frame. The Israel-Darmois conditions allow one to express the jump in the extrinsic curvature as s function of the shell composition. In particular, we get for a static shell at $r=R$~\cite{Israel:1966rt},
\beq
\sigma&=&-\frac{1}{4\pi R}\left(\sqrt{f_+}-\sqrt{f_-}\right)\,,\label{junction1}\\
%
%
P&=&\frac{1}{8\pi R}\left(-4\pi R\sigma+\sqrt{f_+}\frac{R h_+'}{2 h_+}-\sqrt{f_-}\frac{R h_-'}{2 h_-}\right)\,.\label{junction2}
\eeq
The strategy is then to integrate Eqs.~\eqref{eqE1}--\eqref{eqE2} from infinity, with appropriate boundary conditions, inwards to the shell; then use the matching conditions to get across the shell, and finally match onto Schwarzschild interior.

A perturbative analysis is perhaps more illuminating. In the small $Q$ limit the metric ansatz reads
\be
ds^2=-\left(1-\frac{2M}{r}+H\right)dt^2+\frac{1}{1-\frac{2M}{r}+F}dr^2+r^2d\Omega^2\,,\nonumber
\ee
where $H$ and $F$ are radial functions to be determined.
In the interior $\Phi={\rm const}$ and the metric is Schwarzschild with $H=F=C/r$, whereas in the exterior we have
\beq
\Phi'&=&\frac{Q}{r(r-2M)}\,,\\
F&=&\frac{2\pi Q^2}{Mr} \log{\left(\frac{r}{r-2M}\right)}\,,\\
H&=&-\frac{2\pi Q^2}{M^2 r}\left[2M+ (r-M) \log\left(\frac{r-2 M}{r}\right)\right]\,.
\eeq
where we have imposed asymptotic flatness and $M$ is the total mass in the Einstein frame. Note that the latter is different from the internal mass of the Schwarzschild metric, whose horizon is located at $r_h\equiv 2 M_{\rm int}=2M-C$. At large distances $\Phi\sim Q/r$. In the physical Jordan frame, this corresponds to a shell with an effective scalar charge $\propto Q$. Indeed, using Eq.~\eqref{metricEinstein} and the expansion~\eqref{expansionA}, the large-distance expansion of the metric $g_{tt}$ in the Jordan frame reads
\begin{equation}
 -g_{tt}=1-\frac{2 M-2 A_1 Q}{r}+\frac{A_1^2 Q^2-2 A_1 MQ+2 A_2 Q^2}{r^2}.\nn
\end{equation}
Therefore, in theories with $A_1\neq0$ the scalar field contributes to the total physical mass. Furthermore, the spacetime acquires a scalar charge given by the coefficient of the $1/r^2$ term above. When $A_1=0$, the scalar charge is proportional to $\sqrt{A_2}Q$.

For a thin-shell the scalar charge $Q$ is a function of $\sigma, P$ and it is determined by the Klein-Gordon equation,
\be
\frac{d}{dr}\left(r^2\sqrt{hf}\Phi'\right)=-\frac{A'(\Phi)}{A(\Phi)}S\delta(r-R)\,.
\ee
where $S=S_{ab}^E\gamma^{ab}=(2P-\sigma)$ is the trace of the surface stress-tensor in the Einstein frame, specified by $\sigma,P$. 
Integrating across the shell we find
\be
Q=\left.\frac{A'}{A}\right|_{R}(\sigma-2P)\,,
\ee
where $A'$ and $A$ are to be evaluated at the shell's location.
Therefore, for a given coupling specified by $A(\Phi)$, the charge $Q$ is uniquely determined by the thermodynamical properties of the shell and by the value of the scalar field at the radius, $\Phi_R\equiv\Phi(R)$. As expected, $Q=0$ when $A'$ vanishes at the radius, i.e. when $\Phi_R=\Phi^{(0)}$. In this case, the scalar field is constant through the entire spacetime and the solution reduces to its GR counterpart. Finally, for a given $Q$, the junction conditions~\eqref{junction1} and \eqref{junction2} can be solved to get $C$ and $R$ in terms of $\sigma$ and $P$.

To summarize, the problem is well-defined: once the matter content is specified, the equations given above allow one to determine unambiguously 
the metric coefficients and scalar field configuration. In 
Fig.~\ref{fig:shell}, we show the internal mass, $M_{\rm int}\equiv 
M-C/2$, the shell radius $R$ and the Kretschmann scalar 
$K=R_{abcd}R^{abcd}$ at the black-hole radius as functions of the 
scalar charge $Q$ for the nonperturbative solutions obtained solving the exact field equations 
numerically keeping the total mass $M$ fixed. Note that the perturbative solutions agrees very well with the exact ones even if the former are only valid to ${\cal O}(Q^2)$~\cite{letter}.

In 
Fig.~\ref{fig:shell} we have normalized all quantities by their values in GR, i.e. $Q=0$, and the horizontal axis has been rescaled by $Q_{\rm extr}$, which corresponds to the extremal solution such that $M_{\rm int}(Q_{\rm extr})=0$. With these normalizations, the quantities $M_{\rm int}$ and $K$ are almost universal for different values of $P$.
As $Q\to Q_{\rm extr}$, the internal mass decreases until the black-hole radius 
$r_h=2M_{\rm int}$ vanishes and the curvature invariant diverges. For large values of $Q$, the structure of the hairy black 
hole can be very different from its GR counterpart.

We have thus constructed nonlinear, hairy solutions of scalar-tensor theories with a black hole at the center. 
Note that these solutions are specificied by an independent parameter $Q$ or, equivalently, by the constant value of the scalar field in the interior, $\Phi_R$. This is due to the fact that the thin-shell does not backreact with the gravitational sector. In a more realistic configuration, matter is described by a stress-energy tensor with nonvanishing support and the dynamics of the matter fields is governed by the conservation $\nabla^a T_{ab}=0$ in the Jordan frame. Eventually, the asymptotic behavior of the scalar field is determined by the matter configuration near the BH and one expects that regular solutions (at the horizon and at infinity) would require $Q$ to have a specific value. This is indeed the case in spontaneously scalarized neutron star~\cite{Damour:1993hw,Damour:1996ke,Harada:1997mr,Pani:2010vc}. In that case, besides the GR solution, only a finite number of scalarized configurations exist which satisfy both the Klein-Gordon and the modified Tolman-Oppenheimer-Volkoff equations.

We have not shown that scalarized solutions arise as the end-state of the instability of the corresponding GR solutions. However, 
the family of solutions given above are the only static and spherically symmetric solution to Einstein equations with a spherical matter shell, so
they have to be the end state of the instability of a Schwarzschild black hole triggered by spherically symmetric modes.
It would be interesting to follow the nonlinear time-dependency of the instability and the dynamical approach to this kind of nonlinear solutions.

\begin{widetext}
\begin{center}
\begin{figure}[h!]
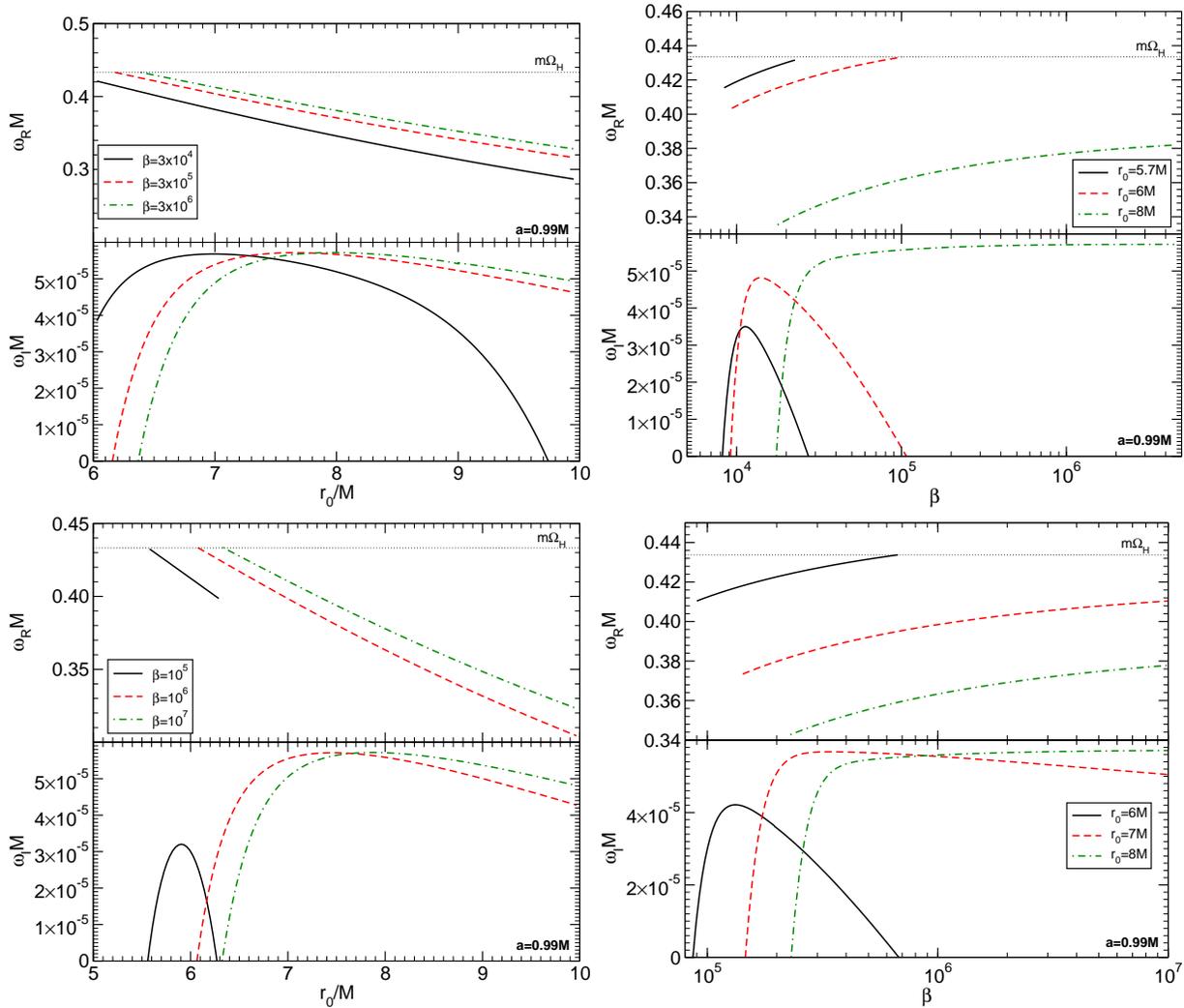

\begin{center}
\begin{tabular}{cc}
\epsfig{file=modes_model_G.eps,width=8cm,angle=0,clip=true}&
\epsfig{file=modes_model_G_beta.eps,width=8cm,angle=0,clip=true}\\
\epsfig{file=modes_model_G_n4.eps,width=8cm,angle=0,clip=true}&
\epsfig{file=modes_model_G_beta_n4.eps,width=8cm,angle=0,clip=true}
\end{tabular}
\end{center}
\caption{\label{fig:modelG}
Summary of $l=m=1$ superradiantly unstable modes for a matter profile characterized by ${\cal G}=\Theta[r - r_0]\beta(r-r_0)/r^n$ with $n=3$ (top panels) and $n=4$ (bottom panels) and various values of $r_0$ (curves are truncated when the modes become stable). For large enough
$\beta$ the system behaves as a black hole bomb \cite{Press:1972zz,Cardoso:2004nk,Dolan:2012yt}, a black hole enclosed in a cavity with radius $r_0$.
Accordingly, the resonant frequencies scale as $1/r_0$ (upper left panel) and for small $r_0$ the instability is quenched (lower left panel).
For large couplings $\beta$, the barrier is high enough that the mechanism is efficient and the instability growth rate is roughly independent
of $\beta$ (c.f. lower right panel. The $r_0=6M$ case is marginal, and increasing $\beta$ has the effect of slight increasing $\omega_R$, sufficient
to quench the instability).
}
\end{figure}
\end{center}
\end{widetext}
%
\section{Spontaneous superradiant instability \label{sec:superradiant}}
When the effective mass squared~\eqref{effectivemass} is positive, spontaneous scalarization as defined before does not occur. However, the appearance of an effective mass
raises the interesting prospect that a ``spontaneous superradiant instability'' is present for rotating black holes. These two instabilities
are different in nature and, in principle, lead to two very distinct end states. The superradiant instability is expected to terminate in a GR solution
with zero scalar field, and lower black hole spin, while spontaneous scalarization gives rise to a nontrivial scalar profile. 
Superradiance requires the presence of an ergosphere whereas spontaneous scalarization does not (and hence can take place even in spherically symmetric spacetimes).

We now show that a spontaneous superradiant instability is also a generic effect of scalar-tensor theories in the presence of matter.
For simplicity, we look for separable solutions of the Klein-Gordon equation (we discuss below potential generalization to non-separable solutions). We found that the following ansatz separates 
and simplifies the treatment significantly, while retaining enough generality:
\beq
\mu_s^2(r,\theta)&=&\mu_0^2+2\frac{{\cal F}(\theta)+{\cal G}(r)}{a^2+2r^2+a^2\cos2\theta}\,,\label{separable}\\
\varphi&=&\Psi(r)S(\theta)e^{-i\omega t+im\phi}\,.
\eeq
The term $\mu_0$ plays the role of the canonical mass term of a massive scalar, whereas $\mu_s$ is the effective mass.
We then get the following coupled system of equations
%
\beq
&&\frac{1}{\sin\theta}\frac{d}{d\theta}\left(\sin\theta\frac{d}{
d\theta}S(\theta)\right)\label{eqang}\\
&&+\left[
a^2\left(\omega^2-\mu_0^2\right)\cos^2\theta-\frac{m^2}{\sin^2\theta}-
{\cal F}+\lambda \right]S(\theta)=0\,,\nn\\
&&\Delta 
\frac{d}{dr}\left(\Delta\frac{d\Psi}{dr}\right)+\left[
\omega^2(r^2+a^2)^2-4aMrm\omega+a^2m^2\right.\nn\\
&&\left.-\Delta\left({\cal 
G}+r^2\mu_0^2+\lambda+a^2\omega^2\right)\right]\Psi=0\,,\label{eqrad}
\eeq
%
where $\lambda$ is a separation constant, to be found imposing regularity conditions on the angular wavefunction $S(\theta)$. Since our purpose here is not to perform an exhaustive analysis of all possible matter profiles but, merely to point out a new mechanism
that turns black hole solutions in GR unstable, we simply discuss different ad-hoc models, all with ${\cal F}=0$:
\beq
i)\quad \mu^2_{0}&\neq& 0\,,\quad {\cal F}={\cal G}=0\,,\nn\\
ii)\quad \mu^2_{0}&=& 0\,,\quad {\cal F}=0\,,{\cal G}=\mu^2r^2\nn\\
%
%
%
iii)\quad \mu^2_{0}&=& 0\,,\quad {\cal F}=0\,,{\cal G}=\beta\Theta[r - r_0](r-r_0)r^{-n}\,.\nn
%
%
\eeq
These matter profiles are not general enough, and further investigation is clearly necessary in order to understand realistic configurations 
such as accretion disks. Such configurations do not separate and therefore do not belong to class \eqref{separable}.
Methods such as those used in Ref.~\cite{Pani:2012vp,Pani:2012bp,Witek:2012tr,Dolan:2012yt} would be required in this case. However, the three families of profiles above capture the most important features of the most relevant spherically symmetric matter distributions, to wit, the first two are approximately constant density cases, while the third one tries to capture halos of matter surrounding a black hole. The last model has an inner boundary, so it can serve as  a first crude approximation to accretion disks, whose inner edge coincides with the last stable circular orbit.

The first two cases can be handled numerically using a highly accurate continued fraction representation of both radial and angular wavefunction,
by a trivial extension of Leaver's method~\cite{Leaver:1985ax,Berti:2009kk,Pani:2012bp}. The first case is identical to Detweiler's setup and it yields the well-known massive scalar field instability \cite{Detweiler:1980uk,Cardoso:2005vk,Dolan:2007mj}. We find that the second profile is also very similar to the massive scalar field case, and yields similar results. We do not consider these any further except to stress that they lead to an instability
whose physical nature is different from the spontaneous scalarization as defined in the previous section.

The third profile was handled by expanding the separation constant $\lambda$ in powers of $a\omega$ \cite{Berti:2005gp}, and using this expansion to directly integrate the radial equation \cite{Berti:2009kk,Pani:2012vp}. We look for the complex eigenvalues:
\be
\omega=\omega_R+i\omega_I\,,
\ee
and focus on unstable modes, i.e. those with $\omega_I>0$, which signals a temporal instability. Results are summarized in Fig.~\ref{fig:modelG} for $n=3$ (top panels) and $n=4$ (bottom panels) and for the fundamental $l=m=1$ modes.

The most important aspect to retain from this model is that the instability is akin to the
original black hole bomb, in which a rotating black hole is surrounded by a perfectly reflecting mirror at $r_0$ \cite{Press:1972zz,Cardoso:2004nk,Dolan:2012yt}:
for small $r_0$ there is no instability, as the natural frequencies of this system, which scale like $1/r_0$, are outside the superradiant
regime $\omega\leq m\Omega_H$, with $\Omega_H$ the black hole angular velocity and $m$ being the azimuthal wave number. In fact, it is clear from Fig.~\ref{fig:modelG} that this is a superradiant phenomenon, as the instability is quenched
as soon as one reaches the critical threshold for superradiance. At fixed large $r_0/M$, and for {\it any} sufficiently large $\beta$, there is an instability 
of roughly constant growth rate. Again, in line with the simpler black-hole bomb system, a critical $\beta$ corresponds to a critical barrier height which is able to reflect radiation back. After this point increasing further $\beta$ is equivalent to increasing further the height of the barrier which has no effect whatsoever on the instability.



%
\begin{center}
\begin{figure}[h]
\begin{center}
\begin{tabular}{cc}
\epsfig{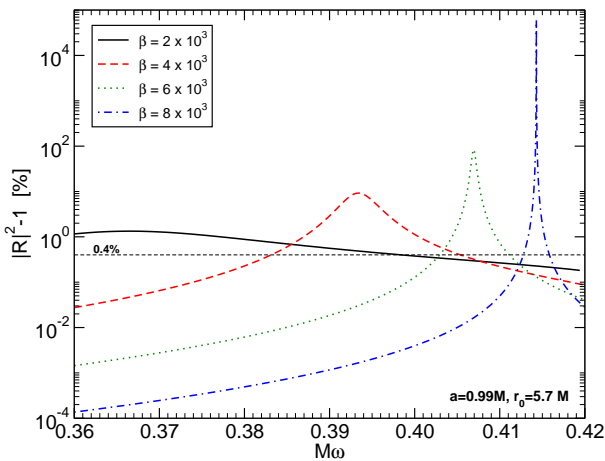}
\end{tabular}
\end{center}
\caption{\label{fig:amplification}
Percentage superradiant amplification factor, $10^2\left(|{\cal R}|^2-1\right)$, for a scalar wave scattered off a Kerr black hole with $a=0.99M$ in a matter profile ${\cal G}=\beta\Theta[r - r_0](r-r_0)/r^3$ as a function of the wave frequency, $\omega$. 
As a reference, the horizontal line corresponds to the maximum superradiant amplification for scalar waves in vacuum, $\approx 0.4\%$.
In this example we set $r_0=5.7 M$, but similar results hold for other choices of $r_0$ and for different matter profiles. The resonances correspond to the excitation of \emph{stable} quasinormal modes.
}
\end{figure}
\end{center}
%

\section{Resonant superradiant amplification}
%

In addition to superradiant instabilities, scalar-matter interactions near spinning black holes give rise to another interesting phenomenon, related to the superradiant scattering of monochromatic waves. 

Because of superradiant energy extraction, an incident wave whose frequency satisfies the condition $\omega<m\Omega_H$ is amplified in a scattering off a spinning black hole~\cite{Teukolsky:1974yv,Press:1972zz}.
Indeed, two asymptotic solutions of Eq.~\eqref{eqrad} (in the near-horizon and near-infinity region, respectively) read
\begin{eqnarray}
 \Psi\sim\left\{ \begin{array}{l}
                  {\cal T}e^{-i(\omega-m\Omega_H)r_*}\qquad r_*\to-\infty \\
                  e^{-i\omega r_*}+{\cal R}e^{i\omega r_*}\qquad r_*\to\infty
                 \end{array}
                 \right.\,, \label{BCsuperradiance}
\end{eqnarray}
where ${\cal T}$ and ${\cal R}$ are the transmission and reflection coefficients, and we have normalized the incident flux to unity.
From the fact that the Wronskian of these two solutions is constant, one immediately obtains the relation $(\omega-m\Omega_H)|{\cal T}|^2=\omega(1-|{\cal R}|^2)$, see e.g. Ref.~\cite{Cardoso:2012zn}. Thus, if the superradiant condition is met one gets $|{\cal R}|^2>1$, i.e. the amplitude of the reflected wave is larger than the incident amplitude.

Within GR superradiant amplification is low. The gain factor, defined as the ratio between the outgoing and ingoing energy fluxes, is at most as large as $0.4\%$ for scattering of scalar waves~\cite{Press:1972zz}. Electromagnetic and gravitational waves are more amplified and the maximum gain factor is about $4\%$ and $138\%$, respectively, for scattering off a nearly-extremal Kerr black hole~\cite{Teukolsky:1974yv}.

We now show that in scalar-tensor theories the presence of matter surrounding a spinning black hole may drastically affect the superradiant amplification of scalar waves. Our setup is standard and consists in solving Eqs.~\eqref{eqang} and~\eqref{eqrad} for a real-frequency scalar wave subjected to the boundary conditions~\eqref{BCsuperradiance}, where we assumed that the matter profile does not have support in the asymptotic regions. The transmission and reflection coefficients can be straightforwardly extracted through a direct integration of the field equations. To improve precision, higher-order series expansion at the horizon and at infinity can be used. The gain factor is then defined as $|{\cal R}|^2-1$.

Our results are shown in Fig.~\ref{fig:amplification} where, for concreteness, we focus on profile iii) listed above, $\mu^2_{0}=0\,, {\cal F}=0\,,{\cal G}=\beta\Theta[r - r_0](r-r_0)/r^3$, and considered $r_0=5.7M$ and a Kerr black hole with $a=0.99M$. Similar results hold for different values of these parameters and for different matter profiles. For small $\beta$, we recover the standard results, with a maximum amplification of~$~0.4\%$ which is displayed in Fig.~\ref{fig:amplification} by a horizontal dashed line.~\cite{Press:1972zz}. On the other hand, as $\beta$ increases, the amplification factor can exceed the standard value by orders of magnitude. This is due to the appearance of spikes at specific frequencies that depend on the parameters of the model. In some cases, the amplification factor can increase by six orders of magnitude or more. Note that $\beta\propto A_2/A_0$ in Eq.~\eqref{effectivemass} and large positive values of $A_2/A_0$ are currently not constrained by observations. Thus, such effect may be used to constrain the parameter space of scalar-tensor theories to unprecedented levels.

The presence of resonances in the amplification factor can be interpreted as follows. At the resonant frequency, $\omega\sim\omega_{\rm res}$, the reflection factor is nearly diverging, ${\cal R}\to\infty$, and this produces a spike in the gain factor. From Eq.~\eqref{BCsuperradiance}, if ${\cal R}\to\infty$ the wave at infinity is (almost) purely outgoing, so that such frequency roughly satisfies the quasinormal mode boundary conditions~\cite{Berti:2009kk}. 

If $\omega_{\rm res}$ were \emph{exactly} a normal frequency of the system, the resonance would have displayed a Dirac delta structure, i.e. it would have had infinity height and zero width. The fact that the resonances shown in Fig.~\ref{fig:amplification} have instead a Breit-Wigner form (with finite height and very narrow width) suggests the existence of very long-lived quasinormal modes with $\omega_R\sim \omega_{\rm res}$ and $\omega_I\ll\omega_R$~\cite{Berti:2009wx}. Such long-lived modes are associated to trapping by potential barriers and they exists in ultracompact stars~\cite{Chandrasekhar:1992ey} and in anti de Sitter black holes~\cite{Berti:2009wx}. They also exist in the case of \emph{massive} scalar perturbations of Kerr black holes~\cite{Detweiler:1980uk}, but in that case the potential well extends to infinity, so that the wave is exponentially suppressed and no superradiant amplification can occur. 

In fact, our previous results for the superradiant instability show that the system also admits normal frequencies, i.e. an eigenfrequency with $\omega_I\equiv0$. For the profile shown in the top panels of Fig.~\ref{fig:modelG}, this happens (for example) when $\beta\approx8194$ and $r_0=5.7M$ and the corresponding frequency is $\omega_R\approx 0.4149/M$. Thus, a wave scattered with frequency $\omega\sim 0.4149/M$ would be infinitely amplified. This would however require an infinitely precise fine-tuning of the frequency for given values of the model parameters. On the other hand, it is clear from Fig.~\ref{fig:modelG} that, in the specific example with $r_0=5.7 M$, modes with $\beta\lesssim 8194$ are stable, so that the resonances shown in Fig.~\ref{fig:amplification} are all associated to the excitation of stable modes.

Scalar-matter interactions 
in scalar-tensor gravity produce an effective scalar mass which can trap long-lived modes and, nonetheless, allows for propagation of scalar waves to infinity. This allows for a new class of long-lived quasinormal modes of Kerr black holes surrounded by matter. Correspondingly to the excitation of these modes, the superradiant gain factor is resonantly amplified.

\section{Conclusions and outlook}
We have shown that black holes surrounded by matter in scalar-tensor theories are generically subjected to two instabilities, which are related to the sign of the effective mass squared of a scalar mode propagating on the black hole background.
Spontaneous scalarization can occur when the effective mass squared is negative and, it is a very generic effect that affects GR solutions when
there is sufficient matter on the outskirts of the event horizon. The spacetime then spontaneously
develops  nontrivial scalar hair supported on the exterior matter profile.
The superradiant instability on the other hand can occur when the effective mass squared is positive and affects rotating black hole solutions. The effectiveness of the instability depends on the matter profile, the spin of the black hole and on the specific scalar-tensor theory considered.

Furthermore, the presence of an effective mass for the scalar field allows for the existence of a new class of long-lived eigenfrequencies of Kerr black holes. Scalar waves whose frequency matches the real part of these quasinormal modes can be resonantly amplified, with amplification factors as large as $10^5$ or more. Similar resonant effects are expected in the flux emitted by binary systems in the extreme-mass ratio limit. It would be interesting to investigate such configurations in scalar-tensor gravity in the presence of matter surrounding the central black hole.

Our results raise a number of questions, two of which would be particularly interesting to understand further:
the dynamical development and final state of these instabilities; and their relevance when it comes to astrophysical black holes and potential observational imprints.
\begin{acknowledgments}
We would like to thank Emanuele Berti, Kostas Kokkotas and Luis Lehner for useful discussions.
V.C. acknowledges partial financial support provided under the European Union's FP7 ERC Starting Grant ``The dynamics of black holes:
testing the limits of Einstein's theory'' grant agreement no.
DyBHo--256667, the NRHEP 295189 FP7-PEOPLE-2011-IRSES Grant, and FCT-Portugal through projects
PTDC/FIS/116625/2010, CERN/FP/116341/2010 and CERN/FP/123593/2011.
P.P. acknowledges financial support provided by the European Community 
through the Intra-European Marie Curie contract aStronGR-2011-298297. 
T.P.S. acknowledges financial support from the European Research Council under the European Union's Seventh Framework Programme (FP7/2007-2013) / ERC Grant Agreement no.~306425 ``Challenging General Relativity'', from 
the Marie Curie Career Integration Grant LIMITSOFGR-2011-TPS Grant Agreement no.~303537, and from the ``Young SISSA Scientist Research project'' scheme 2011-2012.
Research at Perimeter Institute is supported by the Government of Canada 
through Industry Canada and by the Province of Ontario through the Ministry
of Economic Development and Innovation.
\end{acknowledgments}


\end{document}